\documentclass{emulateapj}
\usepackage{psfig}
\usepackage{epsf}



\newcommand{\mum}{$\,\mu$m}
\newcommand{\smm}{$S_{\rm 850 \mu m}$}

\newcommand{\sub}{submillimeter}
\newcommand{\ratio}{S$_{\rm 850 \mu m}$/S$_{\rm 1.4 GHz}$}

\newcommand{\bivar}{$\Phi$(${\cal L}$, ${\cal C}$)}

\newcommand{\ltir}{L$_{\rm TIR}$}

\newcommand{\td}{T$_{\rm d}$}

\newcommand{\sfr}{{\rm\,M_\odot\,yr^{-1}}}

  \def\itm#1 {\vskip10pt \reference{} \square\ {\bf #1} }
  \def\square {\hbox{\vrule width5pt height5pt}}

\lefthead{Lewis, Chapman \& Helou}

\begin{document}
\title{Modeling the evolution of infrared luminous galaxies:
the influence of the Luminosity-Temperature distribution.
}
\author{
G.\,F.\ Lewis,$\!$\altaffilmark{1}
S.\,C.\ Chapman,$\!$\altaffilmark{2}
G.\ Helou,$\!$\altaffilmark{2}
}

\altaffiltext{1}{Institute of Astronomy, 
University of Sydney, Sydney, NSW 2006, Australia;
{\tt gfl@physics.usyd.edu.au}}
\altaffiltext{2}{California Institute of Technology, 
MS 320-47, Pasadena, CA, 91125;
{\tt schapman,ghelou@irastro.caltech.edu}}

\slugcomment{To appear in the Astrophysical Journal}

\begin{abstract}
The evolution  of the luminous infrared galaxy  population is explored
using a pure luminosity evolution model which incorporates the locally
observed luminosity-temperature distribution  for IRAS galaxies.  Pure
luminosity  evolution models  in  a fixed  $\Lambda$CDM cosmology  are
fitted to submillimeter (submm)  and infrared counts, and backgrounds.
It  is  found that  the  differences  between  the locally  determined
bivariate model  and the single  variable luminosity function  (LF) do
not  manifest  themselves  in  the  observed counts,  but  rather  are
primarily apparent in the dust temperatures of sources in flux limited
surveys.   Statistically  significant   differences  in  the  redshift
distributions  are also  observed.   The bivariate  model  is used  to
predict  the  counts,   redshifts  and  temperature  distributions  of
galaxies  detectable by  {\it  Spitzer}.  The  best  fitting model  is
compared  to the  high-redshift submm  galaxy population,  revealing a
median    redshift    for    the    total    submm    population    of
$z=1.8^{+0.9}_{-0.4}$,  in good  agreement  with recent  spectroscopic
studies of  submillimeter galaxies.  The  temperature distribution for
the submm  galaxies is modeled  to predict the radio/submm  indices of
the submm  galaxies, revealing that  submm galaxies exhibit  a broader
spread in  spectral energy distributions  than seen in the  local IRAS
galaxies.
\end{abstract}


\section{Introduction}
\label{secintro}
With an  energy comparable to the  optical/UV background, measurements
of the far infrared background reveal  it to peak at around 200$\mu m$
(Puget   et  al.\   1996;   Fixsen  et   al.\   1998),  arising   from
dust-reprocessing of high energy  radiation and star formation and AGN
activity  in  $z>0$  galaxies.    Such  obscuration  could  be  hiding
approximately  half of the  massive star  formation activity  over the
history  of  the Universe  (e.g.~Blain  et  al.\  1999a). Clearly,  to
unravel  the cosmic  history of  star formation  the evolution  of the
infrared galaxy population needs to understood.

Studies  of the  evolving infrared  galaxy populations  have typically
assumed a small range of template galaxy spectral energy distributions
(SEDs), or even a single SED.  However, infrared galaxies span a large
range   in   properties,    with   60\mum/100\mum\   colors   spanning
$\sim0.4$\,dex  for a  given  luminosity (Soifer  \& Neugebauer  1991,
Chapman et al.\  2003a).  This distribution in dust  SEDs implies that
substantial numbers of both  extremely luminous, yet cold galaxies, as
well as low luminosity, hot galaxies are found.

In  a previous  paper  (Chapman et  al.\  2003a --  hereafter C03)  an
evolving distribution  bivariate in luminosity and  color, \bivar, was
presented, providing consistency with  the broad distribution of submm
galaxies  observed  locally.   This  earlier paper  demonstrated  that
flux-limited  surveys in  various infrared  and  submillimeter (submm)
bands would  subsume a non-negligible fraction of  both cold, luminous
galaxies  and  hot,  faint  galaxies,  the predictions  borne  out  in
observations of  low and moderate  redshift IRAS galaxies, as  well as
microJansky radio  sources.  The overall conclusion of  this study was
that  surveys which select  objects at  either the  cold Raleigh-Jeans
tail of the dust SED, or the hot Wien tail, will preferentially detect
appropriately cold  or hot  objects for a  given luminosity  class, in
much  larger numbers  than expected  if the  temperature distributions
were not taken into account.

\begin{figure}
\centerline{    \psfig{file=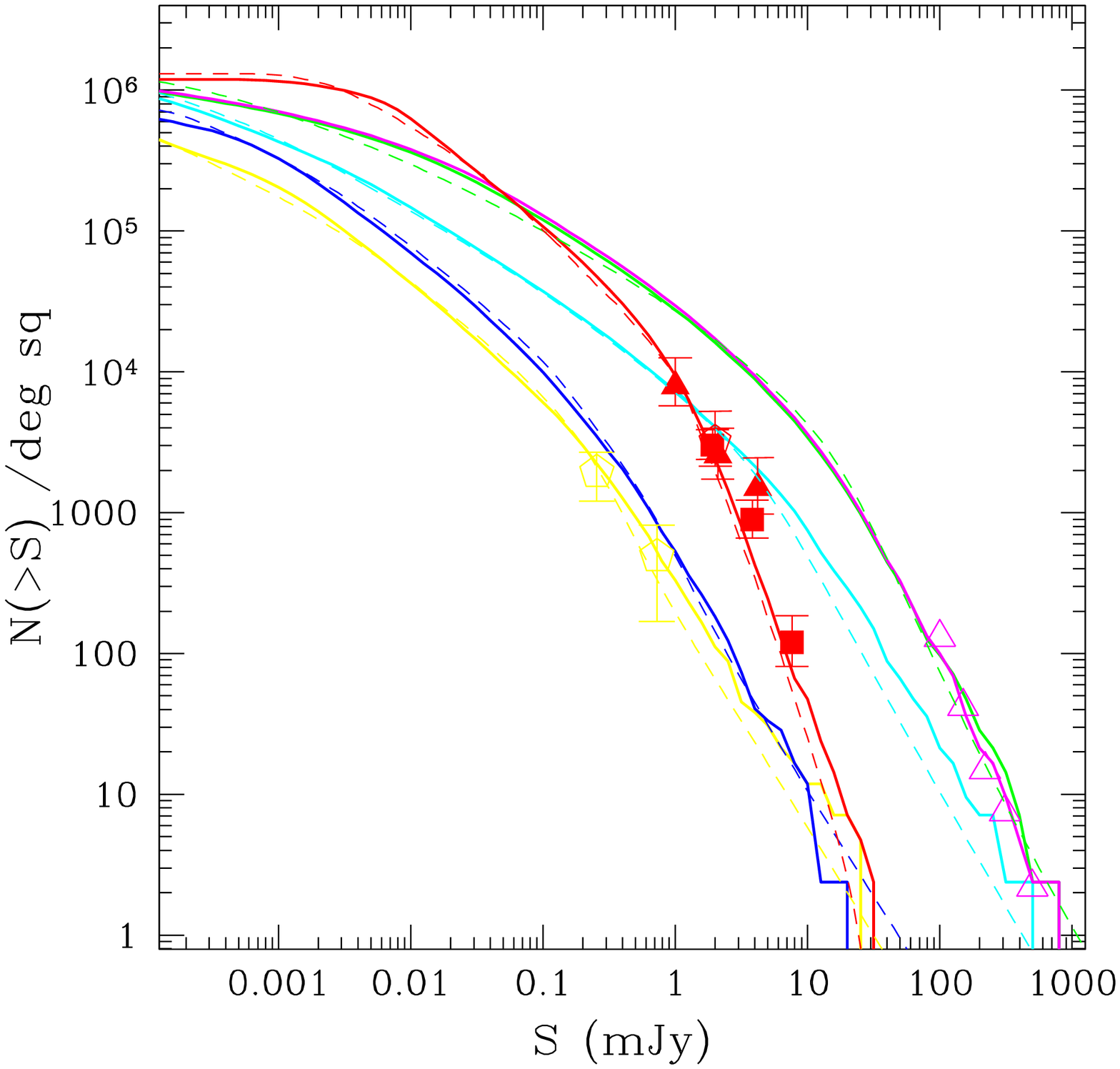,width=3.2in}    }    \centerline{
\psfig{file=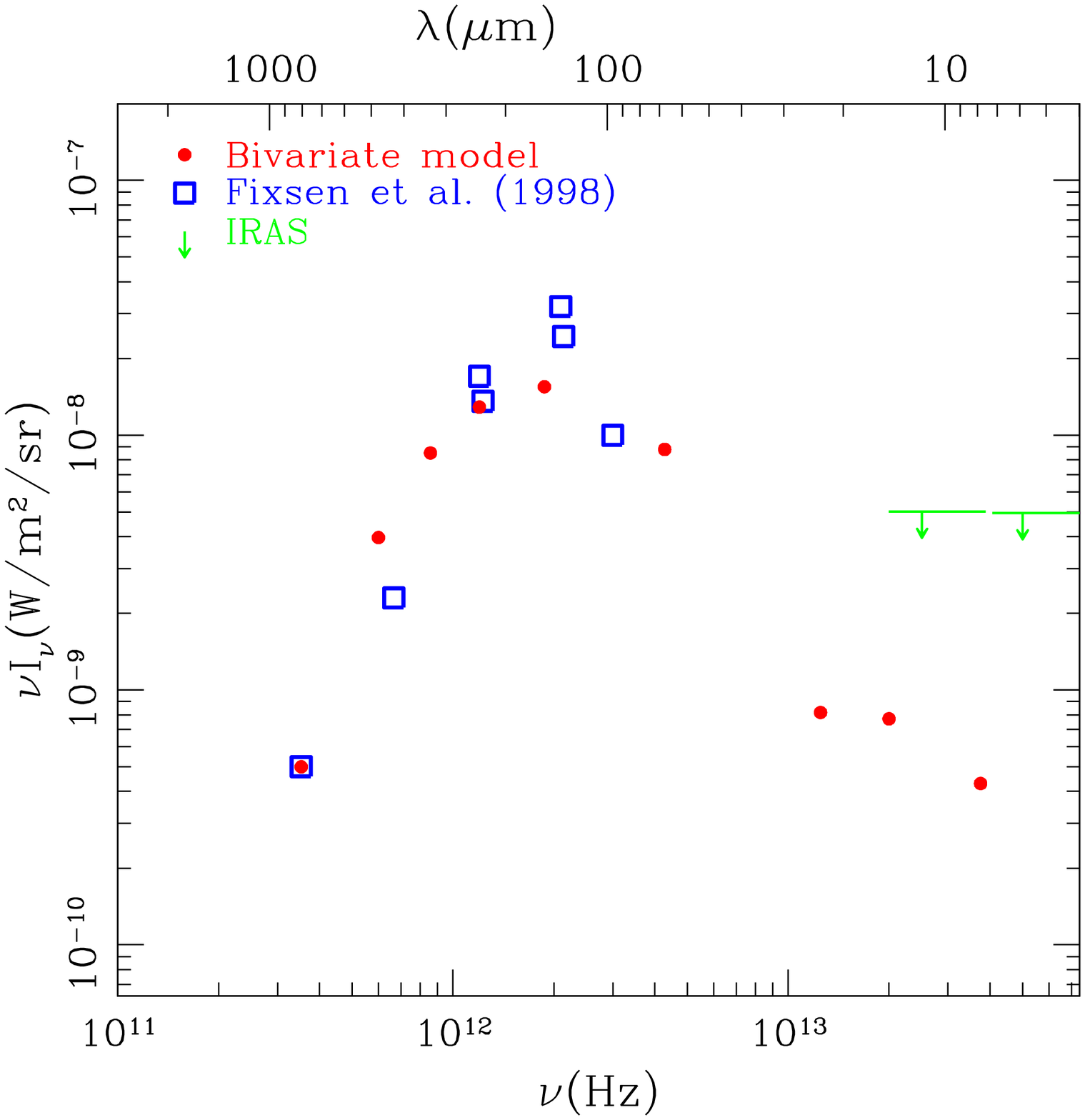,width=3.2in} }
\caption{The upper  panel presents the bivariate model  count from our
best  fit simulation  at 850\mum\  (red), 170\mum\  (purple), 160\mum\
(green), 70\mum\ (cyan), 24\mum\ (blue), and 15\mum (yellow), compared
to SCUBA and ISO counts from the literature (points).  For comparison,
we  overlay the  model counts  of Lagache,  Dole \&  Puget\  (2003) as
dashed lines.  In  the lower panel, the integrated  bivariate model is
compared to the Cosmic Infrared-Background of Fixen et al. (1998). The
upper-limits  at 12$\mu  m$  and 24$\mu  m$  are drawn  from Scott  et
al. (2001).
\label{counts} }
\end{figure}

Submm-luminous,  extragalactic  sources  (Smail, Ivison,  Blain  1997)
currently  provide  our  only  means  of studying  the  high  redshift
infrared galaxy population (Chapman et al.\ 2003b). These galaxies are
now routinely detected with the SCUBA/JCMT and MAMBO/IRAM instruments,
and over 200 blank field sources are now cataloged from cluster lensed
surveys (see Blain et al.\ 2002 for a summary).  The tight correlation
observed locally  between thermal  FIR emission and  synchrotron radio
emission (Helou et al.~1985, Condon 1992) allows the identification of
submm sources, recovering $\sim$65\% of the blank field counts (Ivison
et al.\  2002; Chapman et  al.\ 2003a).  With  spectroscopic redshifts
for the submm  galaxies (SMGs), the possibility exists  to assess more
subtle effects in  our model, and to understand  whether our model can
explain  the range  in properties  subsumed by  the SMGs,  as  well as
explore the  implication of  various selection approaches  employed in
submm surveys.

In  this  paper, the  bivariate  model is  examined  in  light of  the
multi-wavelength counts  and backgrounds, using  a parameterization of
pure  luminosity  evolution.   The  model  is first  used  to  explore
predictions for  the {\it Spitzer Space  Telescope} galaxy populations
in terms of their redshift, color, and temperature distributions.  The
model is  then employed to  better understand the range  in properties
and  redshift  distributions  sampled  by the  radio-identified  submm
galaxy  population,   and  the  importance   of  the  luminosity--dust
temperature  distribution, ${\cal L  - T}$,  and by  extrapolation the
properties  of  the entire  submm  population.   All calculations  are
undertaken in a $\Omega_0=0.3$, $\Lambda_0=0.7$, h=0.65 cosmology.

%
%
\begin{figure*} 
\centerline{               
\psfig{file=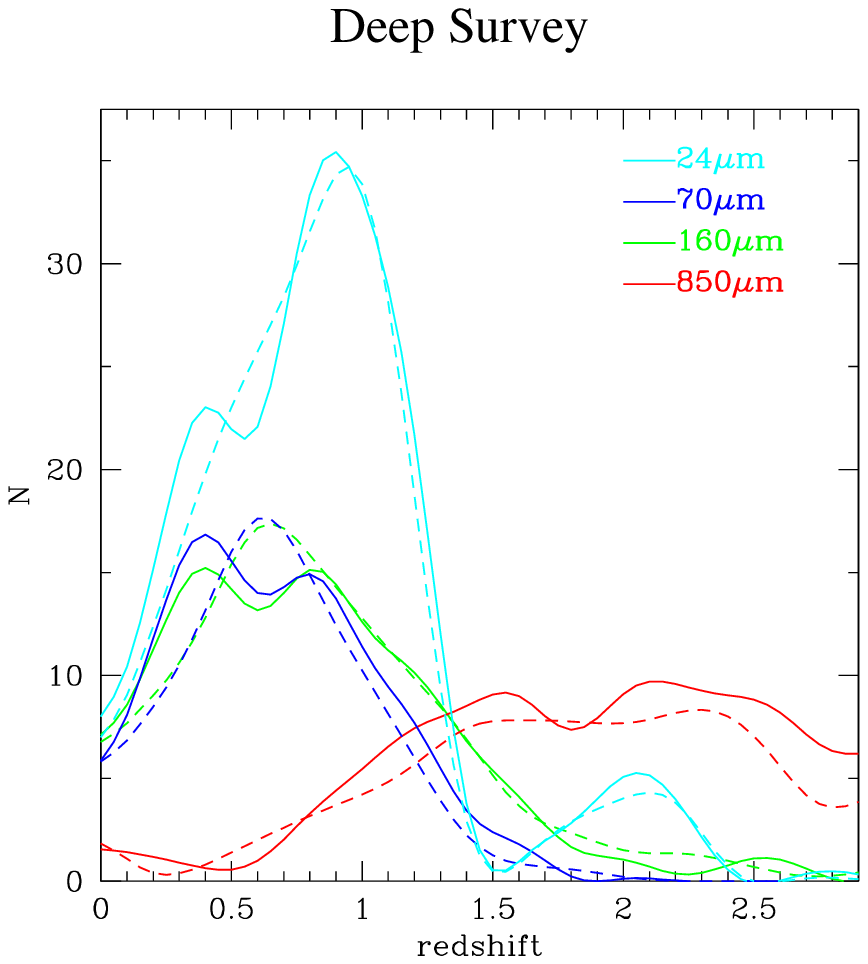,angle=270,width=3.4in}
\psfig{file=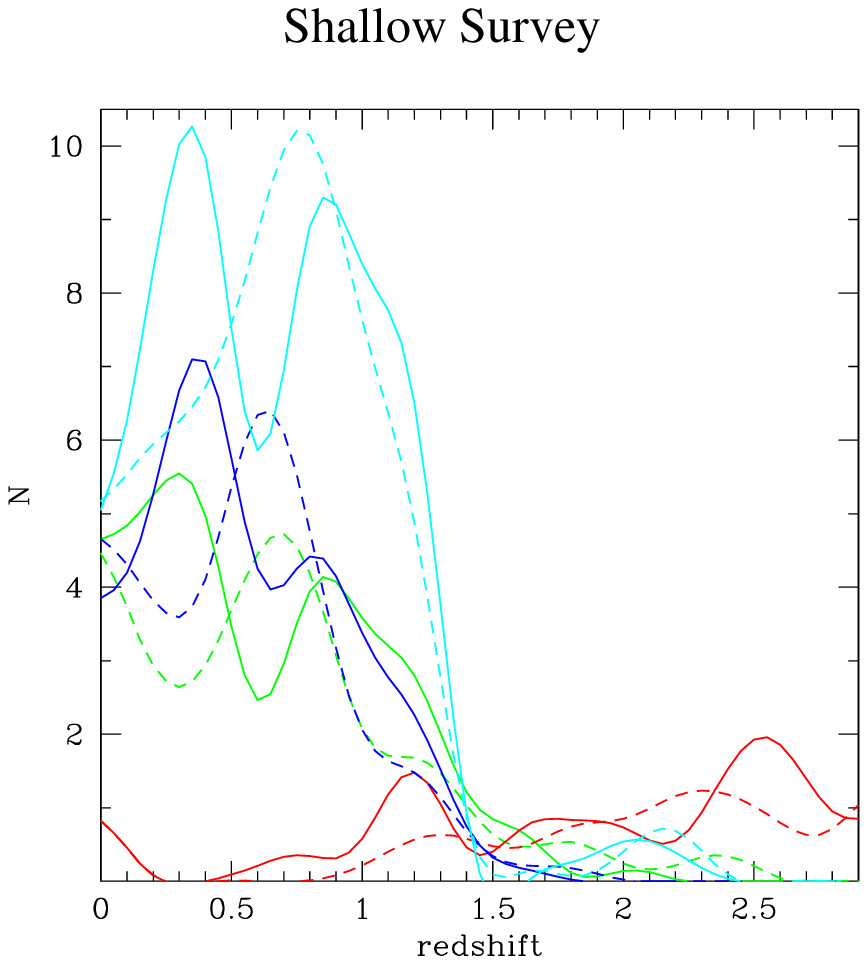,angle=270,width=3.4in} }
\caption{  Redshift  distributions  at  24\mum, 70\mum,  160\mum,  and
850\mum.  The  distributions are  shown for both  a deep (left)  and a
shallow (right)  survey, in  the bivariate  (solid line)  and single
variable  (dashed   line)  models.    The  effect  of   the  bivariate
distribution is to increase the  numbers of both high and low-redshift
sources relative to the single  variable model, due to the temperature
spread.  The shallow survey is  comparable to the expected FLS survey,
with  the 850\mum\ survey  being typical  of current  wide-field SCUBA
surveys. The deep  survey is scaled at all wavelengths  by a factor of
5. \label{nzsirtf} } \addtolength{\baselineskip}{10pt}
\end{figure*}

\section{Evolution of the bivariate LF}
The  evolutionary  model  is  anchored  to the  local  FIR  luminosity
function (LF),  constructed from the  1.2\,Jy sample of  IRAS galaxies
(Fisher et  al.~1995, C03).  Recent work has  emphasized the variation
in dust temperature  found in ULIRGs with T$_{\rm d}$  as low as 25\,K
(Chapman et  al.\ 2002c).  The adopted  form of the  LF represents the
distribution  in  dust  temperatures  found  in  the  1.2\,Jy  sample,
parametrized by the 60\mum/100\mum\  color (a full characterization of
this LF can be found in C03).

The local bivariate LF is then evolved using pure luminosity evolution
with redshift.  While  a range of functional forms  have been employed
in the  literature in  the study  of the various  IR, submm  and radio
populations,  Blain et  al.\ (1999a,b)  have demonstrated  that models
which  are not  strongly  peaked, in  particular  those which  flatten
beyond  a certain  redshift, have  problems over-predicting  the submm
background.   While  it is  clear  that it  will  not  be possible  to
distinguish minute  details of evolutionary  form, this study  aims to
constrain its gross properties.

%
%
\begin{figure*} 
\centerline{
\psfig{file=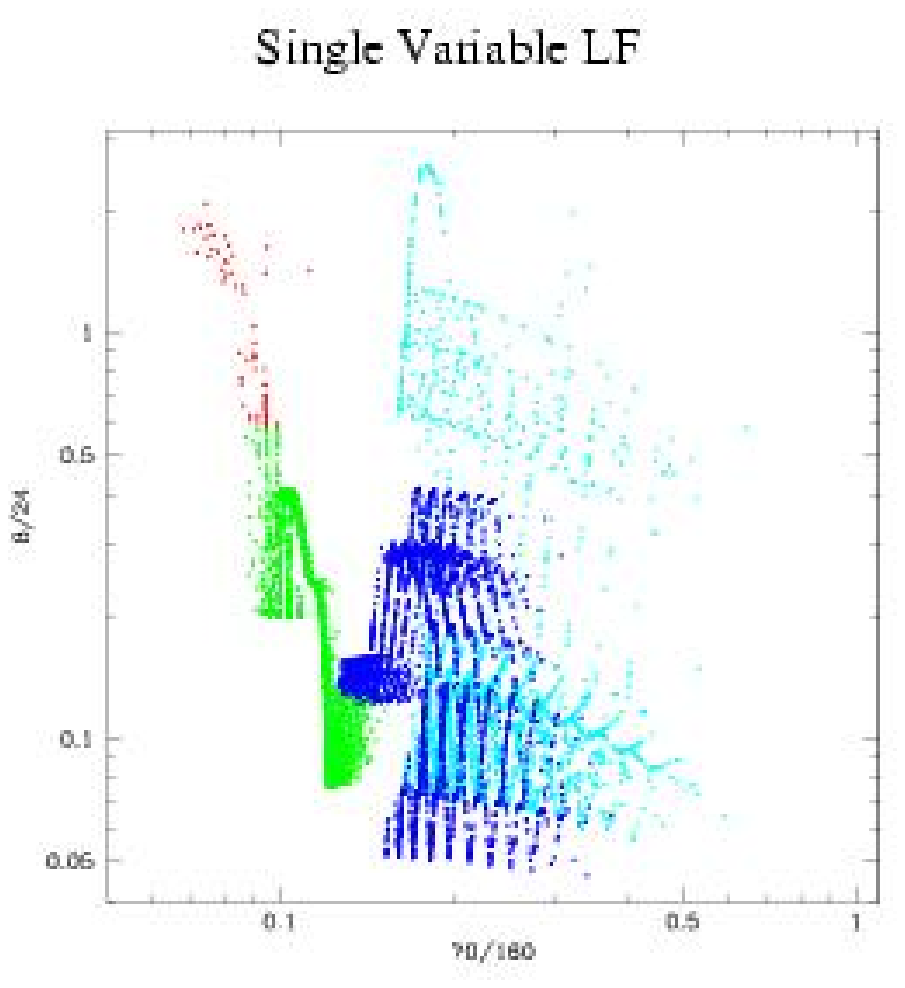,angle=270,width=3.1in}
\psfig{file=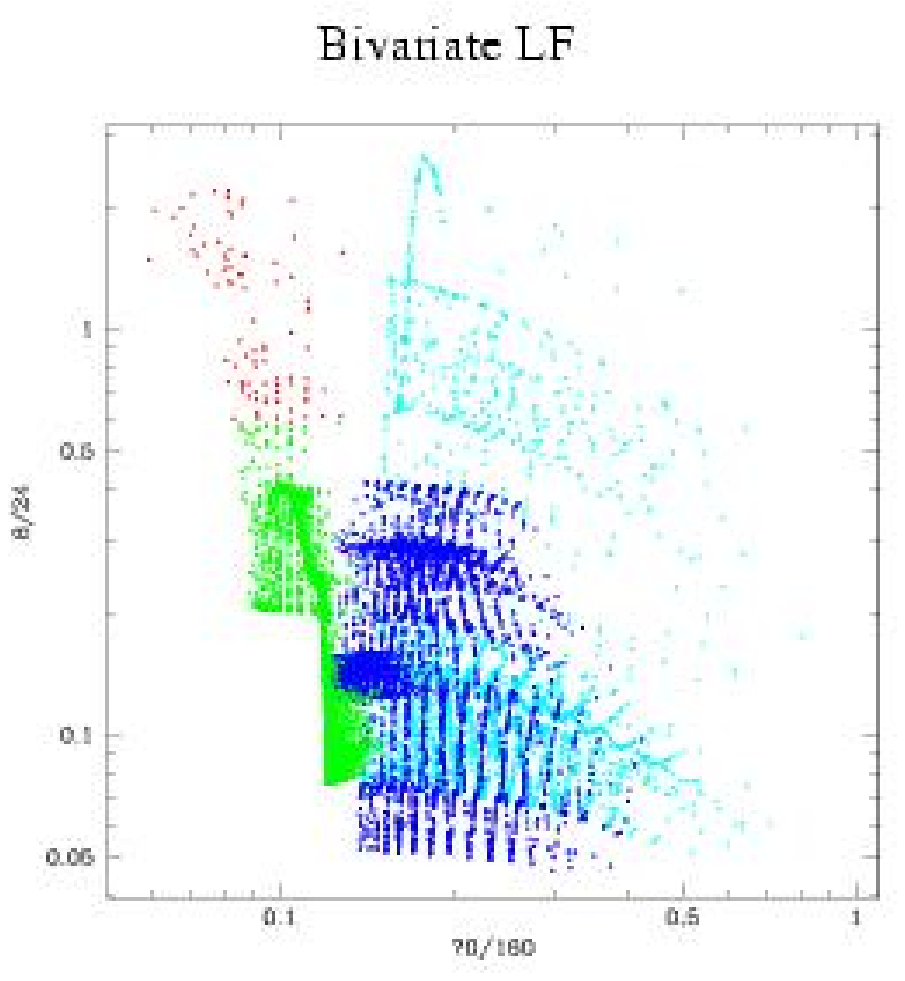,angle=270,width=3.1in}
}
\caption{Color-color diagrams  of galaxies  in a 24\mum\  flux limited
survey (S$_{24  \mu m}>0.2$\,mJy). The left-hand  plot presents $70\mu
m/160\mu m$ versus  $8\mu m/24\mu m$ for the  univariate case, whereas
the right-hand  panel presents the same distribution  for the bivariate
case.  Objects  are color-coded  by redshift: $0<z<1$  (cyan), $1<z<2$
(blue), $2<z<3$ (green), $z>3$ (red).  \label{figcc} }
\end{figure*}

\subsection{Fitting the multi-wavelength counts and backgrounds}
Studies invoking a pure density evolution of the FIR galaxy population
have been shown to grossly over predict the cosmic infrared background
and are  hence unphysical. Other authors have  considered both density
and   luminosity  evolution  of   the  FIR   population,  successfully
recovering  both the  number counts  and backgrounds  (Franceschini et
al. 2001; Chary \& Elbaz 2001;  Xu et al. 2004).  However, it has been
shown  that these  observational  properties can  be explained  within
models adopting  only luminosity  evolution (Blain 1999a).   Given its
simplicity  (with  reduced  parameters),   we  adopt  a  similar  pure
luminosity evolution to examine the cosmic history of FIR galaxies.

The evolution is modeled in a simple form; $\Phi (L,{\cal C}) = \Phi_0
(L/g(z),  {\cal C'})$.  The  color term  of our  LF (${\cal  C'}$ rest
frame  60\mum/100\mum\  flux  ratio)   does  not  evolve,  but  rather
continues  to scale  with FIR  luminosity as  found locally.   C03 and
Blain, Barnard \&  Chapman (2003) have demonstrated that  this lack of
significant 60\mum/100\mum\  color evolution appears to  hold at least
out to moderate  redshifts ($0.3<z<0.9$) in a sample  of IRAS detected
ULIRGs of Stanford et al.\ (2000).  As  the form of the ${\cal L - T}$
distribution  remains  fixed for  all  redshifts,  the only  remaining
parameter  is $g(z)$.  It  should be  noted that  Dunne et  al. (2000)
suggested that the IRAS  galaxy luminosity function may be incomplete,
missing cold galaxies. Recently, this conclusion has been strengthened
by observations by Klaas et al.  (2001) and Bendo et al. (2002; 2004).
Given  that  the current  study  is tied  to  IRAS,  this effect  will
accentuate  the features  of cold  galaxies presented  in  this paper,
although the degree of  IRAS underrepresentation of cold galaxies must
be quantified to fully explore this.

Adopting a Monte Carlo  approach, luminous infrared galaxies are drawn
randomly  from  the evolving  distribution  function.   A galaxy  thus
selected from this model Universe is then assigned a template spectral
energy  distribution from  the catalog  of Dale  et  al.~(2001, 2002),
parameterized by the 60\mum/100\mum\  color, and normalized to the FIR
luminosity.  This  model scenario  does not incorporate  the intrinsic
scatter  in  the FIR/radio  correlation  (0.2\,dex),  as  was done  in
Chapman et al. (2002b;  hereafter C02). Rather this study concentrates
on the properties of the intrinsic dust temperature distribution which
locally  show   a  larger   scatter  (0.3\,dex)  than   the  FIR/radio
correlation,  and should  be expected  to dominate  the  high redshift
radio and submm properties.

C02 adopted an evolution function  with a simple power law peak, $g(z)
\propto (1+z)^{4}$ out to a break redshift, and dropping thereafter as
$g(z)  \propto  (1+z)^{-4}$,  with  no  discontinuity  in  the  zeroth
derivative.   This  power-law  index  was chosen  provided  reasonable
descriptions of  submm and  radio data, allowing  a fit to  the sub-mm
counts  and background (C02).   This evolution  form coupled  with the
bivariate  model  provided  a  reasonable  fit over  the  submm  fluxes
represented by our data (5$<$S$_{850 \mu m}$$<$15\,mJy).  However, the
model  produced too  many  bright submm  sources,  with an  increasing
excess of $>$20\,mJy sources  with increasing peak redshift.  This toy
model cannot  be considered physical,  and the culprit for  the excess
bright sources is found to be in the peak turnover in $g(z)$ producing
a spike of luminous sources  at the tip of our color-magnitude diagram
(CMD)  coupled with  the range  of  T$_{\rm d}$  for each  luminosity.
While this  peak model was  useful for illustrating  various selection
effects  (as  in C03),  it  must be  replaced  with  a more  realistic
evolution form  to provide  a more accurate  description of  the submm
population  and,  therefore, the  functional  form  of  Blain et  al.\
(1999b) is adopted $g(z) =  (1+z)^{3/2} sech^2[{\rm b} \ln(1+z) - {\rm
c}] cosh^2{\rm c}$.

%
%
\begin{figure*} 
\centerline{                  
\psfig{file=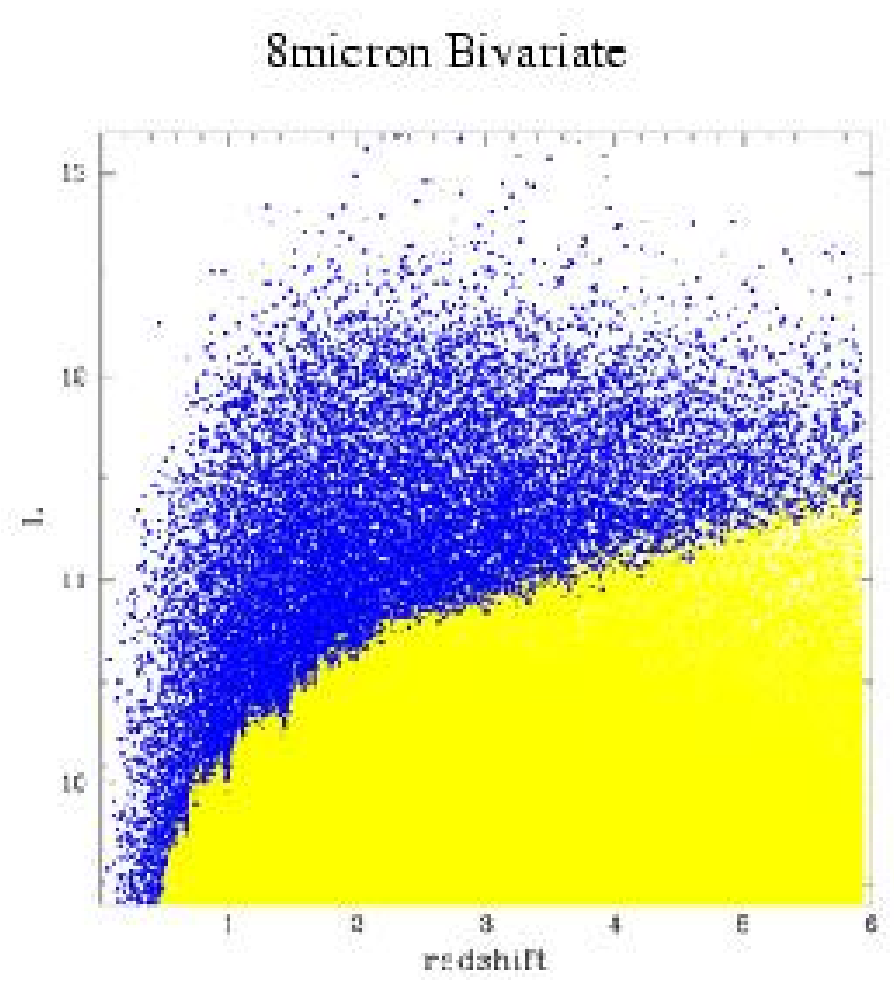,angle=270,width=1.4in}
\psfig{file=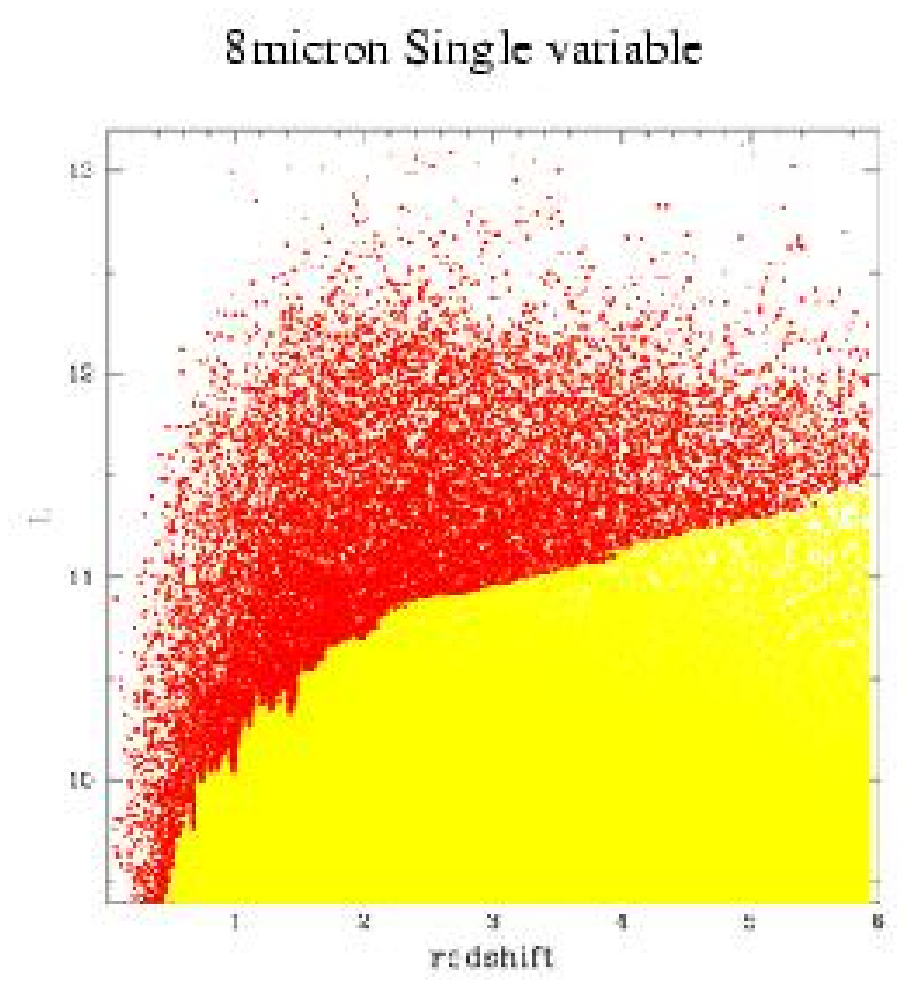,angle=270,width=1.4in}
\psfig{file=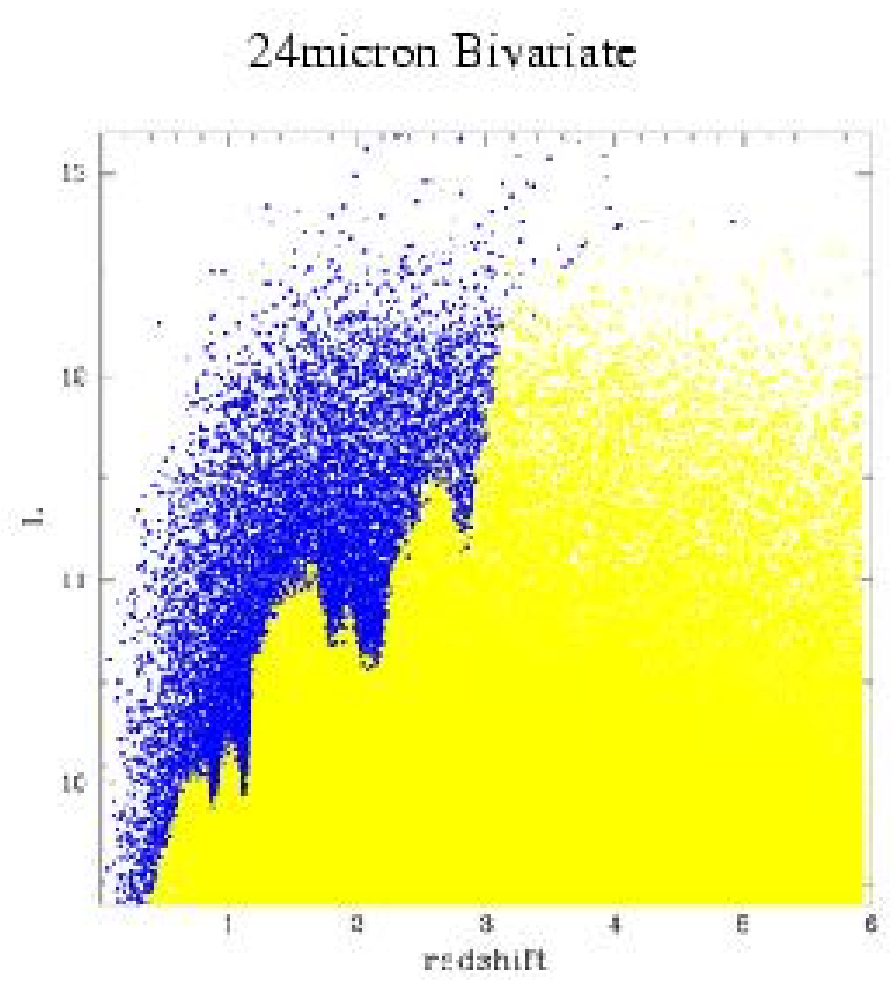,angle=270,width=1.4in}
\psfig{file=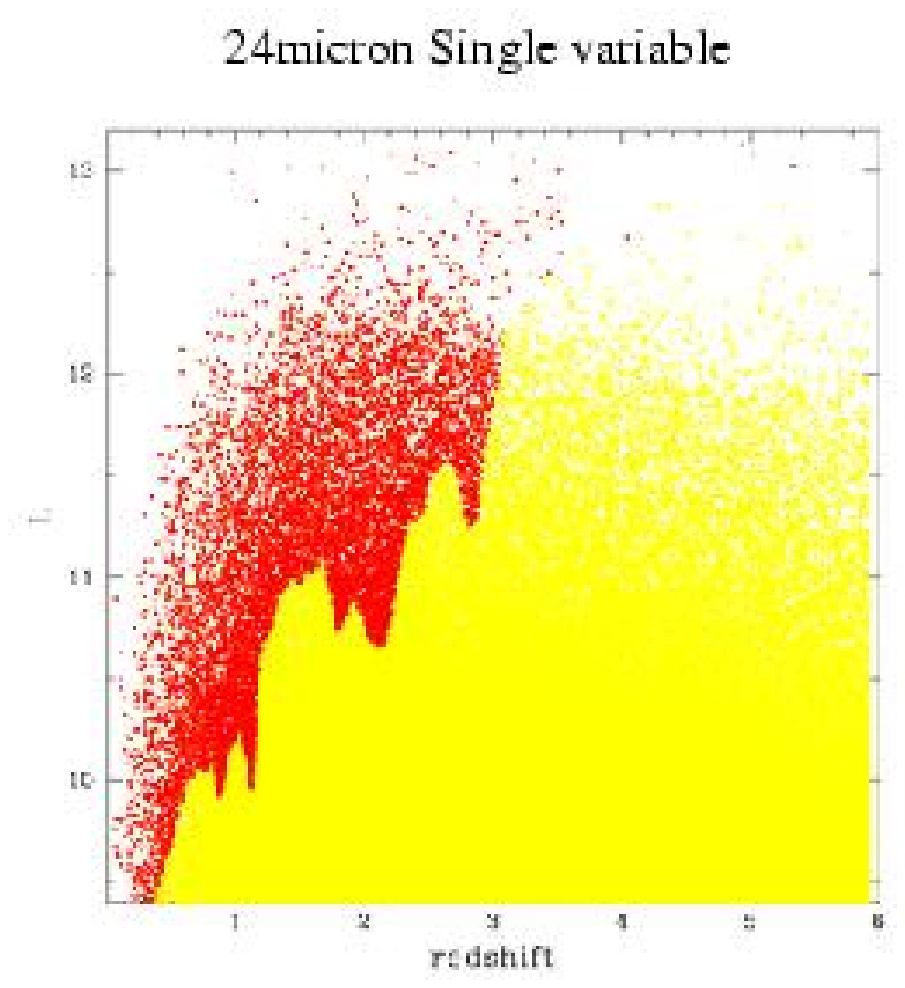,angle=270,width=1.4in}          }         \centerline{
\psfig{file=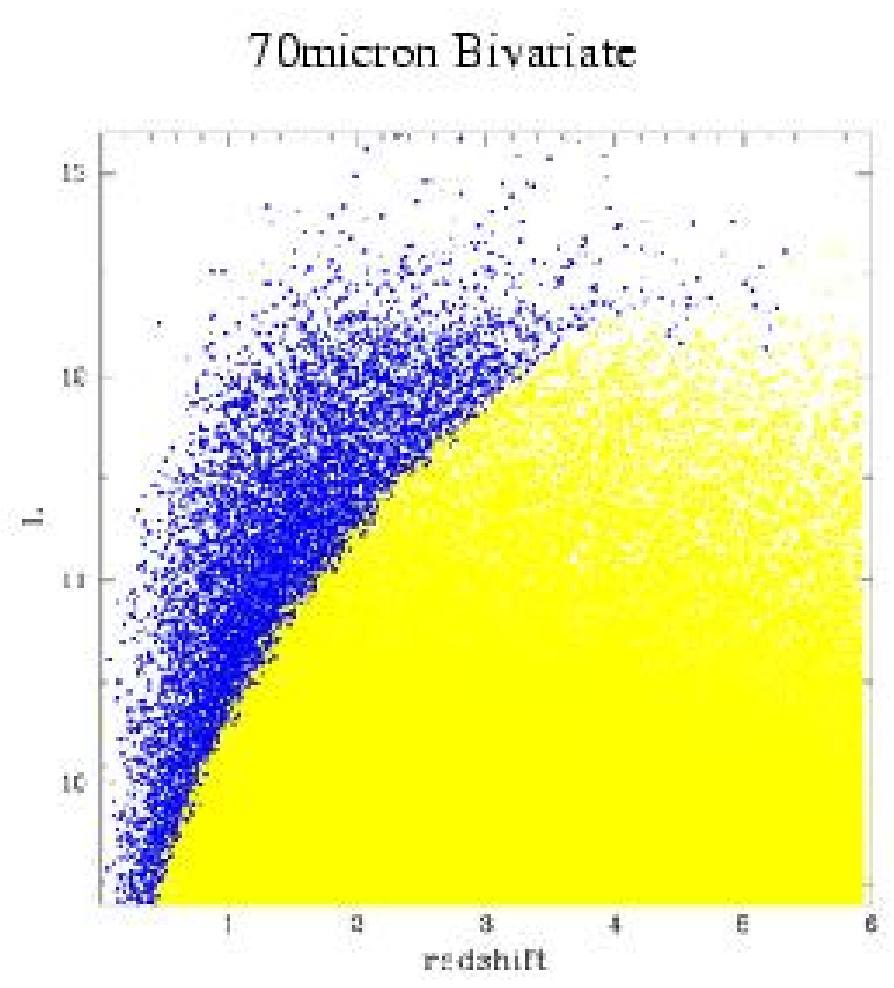,angle=270,width=1.4in}
\psfig{file=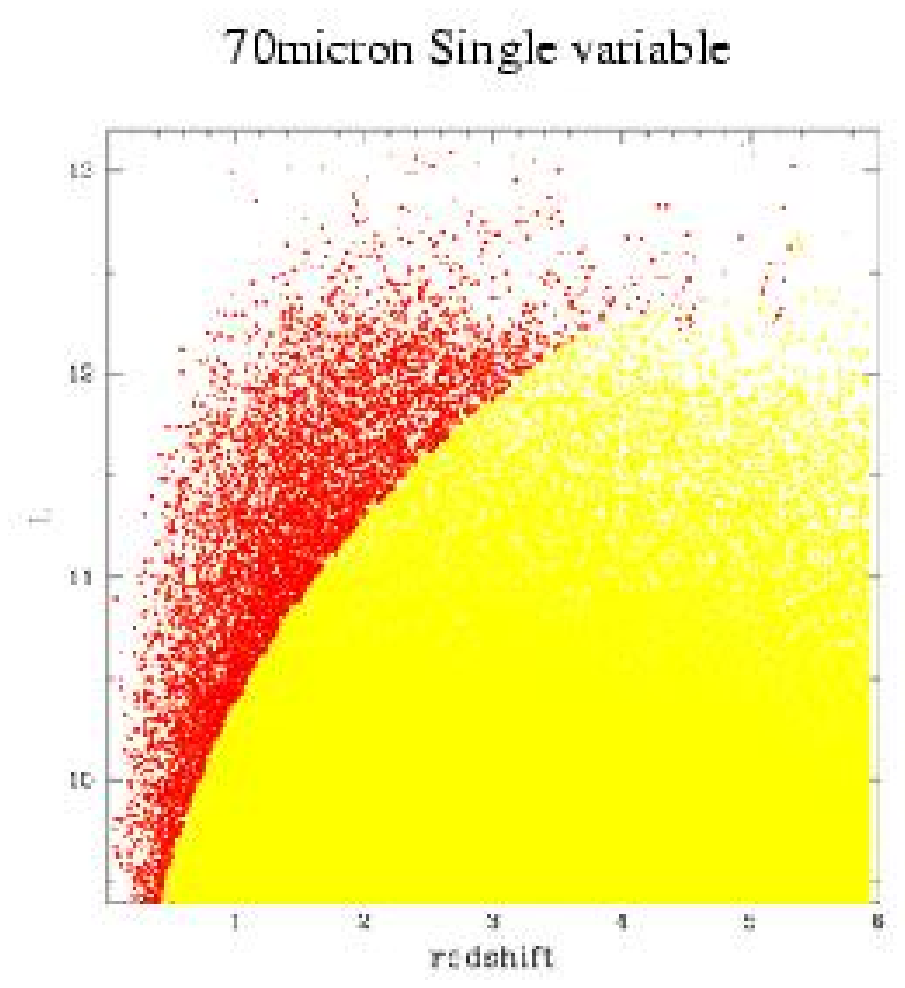,angle=270,width=1.4in}
\psfig{file=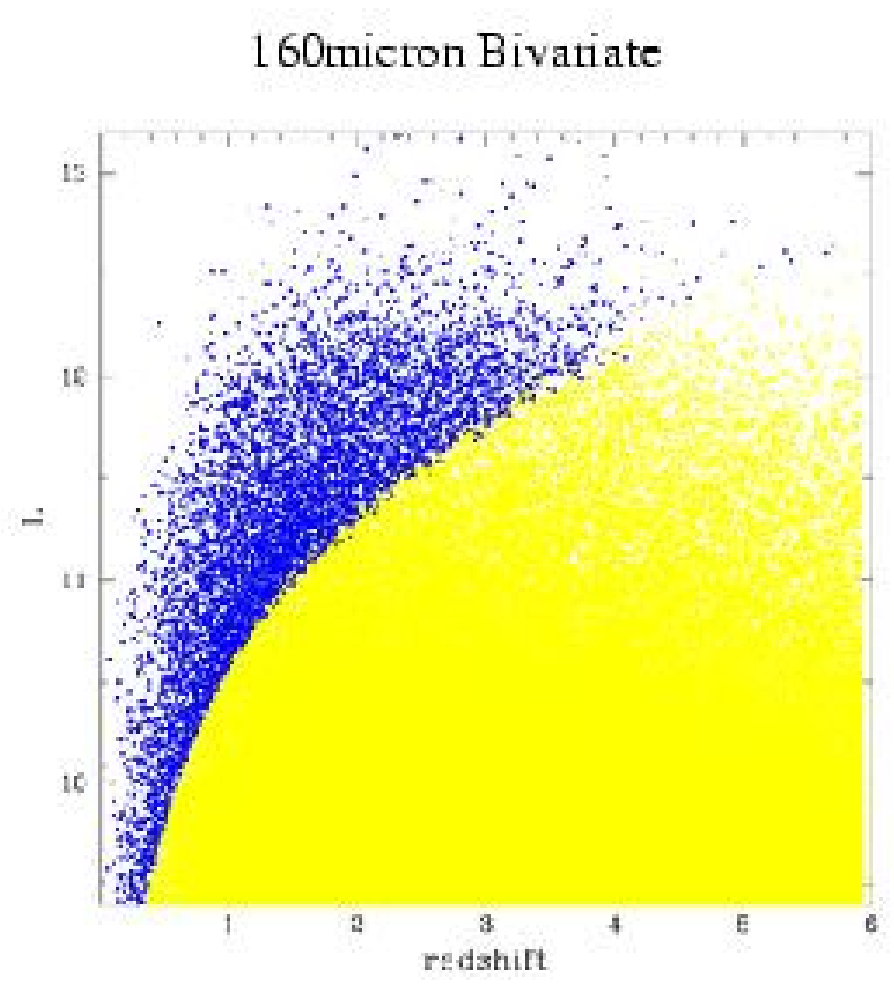,angle=270,width=1.4in}
\psfig{file=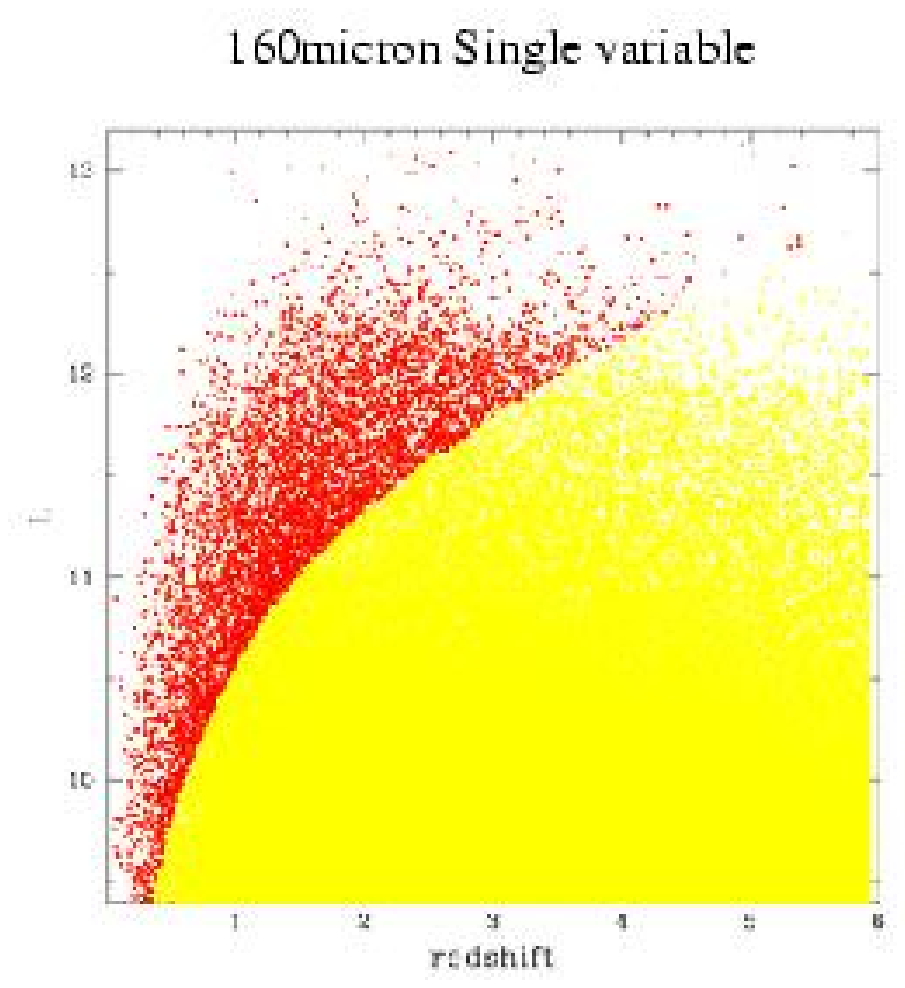,angle=270,width=1.4in} }
\caption{ Comparison of our Monte Carlo representation of the evolving
\bivar\ (blue points) to an  evolving univariate model LF (red points)
with  a  one-to-one  mapping   of  luminosity  to  color.  Each  point
represents  only luminosity and  thus the  distributions are  equal in
both  cases.  The  difference  between models  is  apparent when  flux
limited  surveys are  defined on  the model  points.  We  overlay flux
limited  samplings of  the model  points with  darker symbols.  In the
single  variable LF  model, luminosities  map uniquely  to  fluxes and
lines can be drawn which reflect fixed temperatures.  The upper panels
are for 8\mum\ and 24\mum, while  the lower panels are for 70\mum\ and
160\mum. \label{lz}}
\end{figure*}

In  constraining  the  form  of  the  above  evolution  model,  galaxy
submm/radio  distributions with  varying values  of $b$  and  $c$ were
constructed and  compared to the  differential number count  models at
850$\mu$m  (Barger, Cowie  \&  Sanders 1999)  and  15$\mu$m (Elbaz  et
al. 1999).   For particular set of  values of $b$ and  $c$, the number
counts are  scaled to minimize  the residuals between the  Monte Carlo
simulation and  the modeled  number counts. For  the purposes  of this
study, it was assumed that  the models possessed an uncertainty of 0.2
dex   and   the  resulting   best   values   of   $b$  and   $c$   are
$2.10^{+0.50}_{-0.60}$  and $1.81^{+0.24}_{-0.46}$  respectively.  The
range of parameters  effectively shift the $z_{\rm peak}$  of the peak
evolution function such that $z_{\rm peak} = 1.8^{+0.9}_{-0.4}$.

Bivariate  model  count  from  our  best fit  simulation  at  850\mum,
170\mum, 160\mum, 70\mum,  24\mum, and 15\mum, are shown  in the upper
panel of Fig.~\ref{counts}. The  models are overlaid on measured SCUBA
(850\mum --  Blain et  al.\ 2002)  and ISO (15\mum\  -- Elbaz  et al.\
1999; and  170\mum -- Dole et  al.\ 2001) counts  from the literature.
For comparison, we overlay the model  counts of Dole et al.\ (2003) as
dashed lines,  which agree  remarkably well with  our models.   At the
brightest end of the counts, our model is constrained by the length of
our Monte Carlo runs, and we are dominated by small number statistics.
In  the  lower  panel   of  Fig.~\ref{counts}  presents  compares  the
integrated  flux in  the  bivariate model,  compared  to the  observed
Cosmic Infrared Background (CIB; Fixen  et al.  1998); again, the best
fit model from the counts accounts for the observed CIB distribution.

Previous modeling (C02) found a reasonable fit to the data with a peak
evolution redshift of $z=3$, when all objects were assumed to have hot
dust temperatures  ($\sim50$\,K), similar to the  local ULIRG, Arp220.
The  current  model ties  the  dust temperature  of  a  source to  its
luminosity as  observed locally for the IRAS  1.2\,Jy sample galaxies,
whereby  a galaxy with  a FIR  luminosity of  10$^{12}$\,L$_\odot$ has
T$_{\rm  d}$=35\,K and  dust emissivity  $\beta=1.6$, as  described in
Dale  \&  Helou  (2002).    In  addition,  the  60\mum/100\mum\  color
distribution  observed  locally  is  adopted  in  the  FIR  luminosity
function,  resulting in  a tail  of cold  luminous galaxies  which are
preferentially  selected with  SCUBA at  850\mum\ (e.g.,  Blain 1999b,
Eales et al.~1999, Chapman et  al.\ 2002c).  As expected (see also the
discussion in  C02), the best-fitting peak evolution  redshift must be
lower  if a  population  of  luminous colder  sources  exists at  high
redshifts.

%
%
\begin{figure*} 
\centerline{
\psfig{file=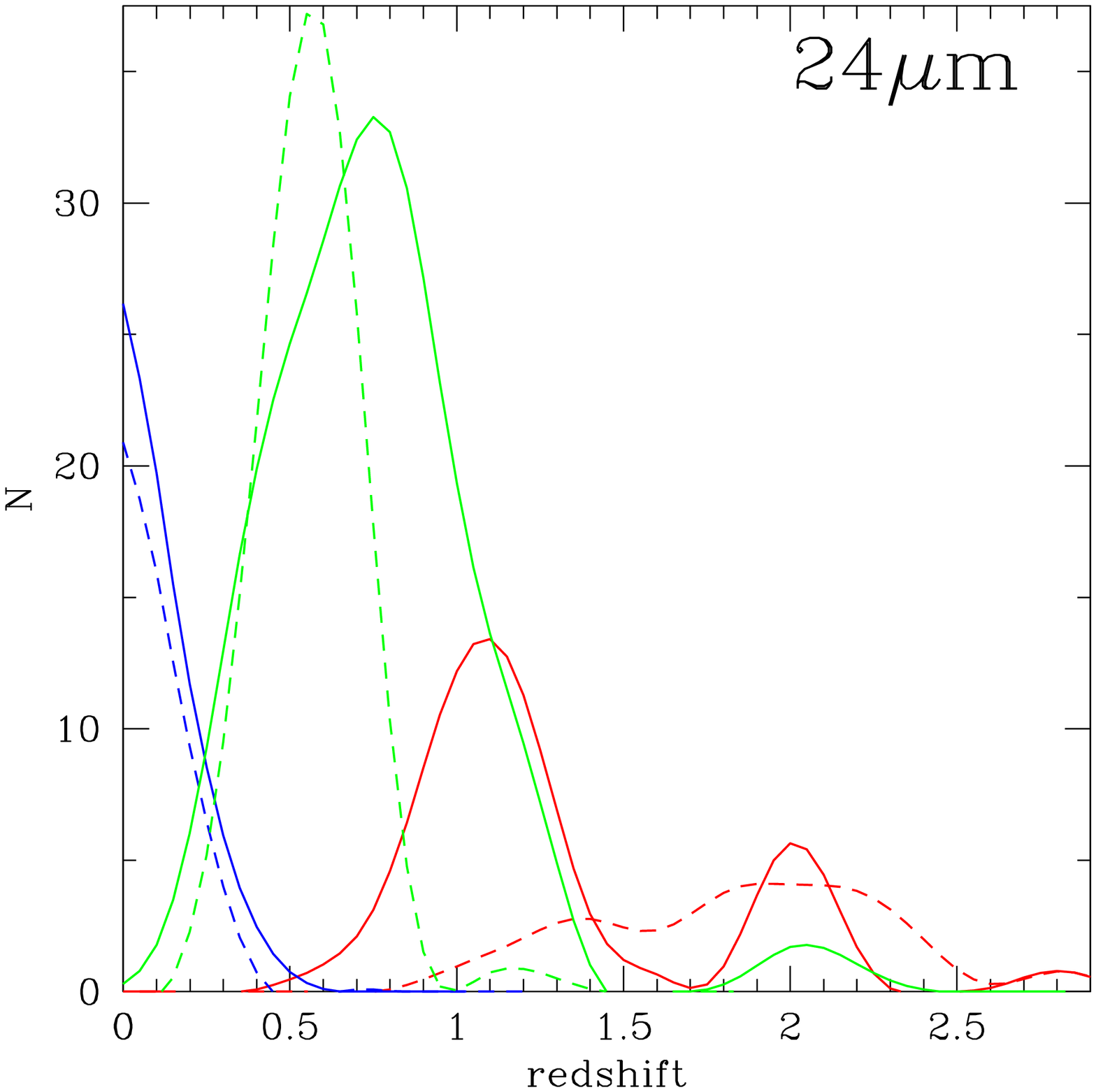,width=1.4in}
\psfig{file=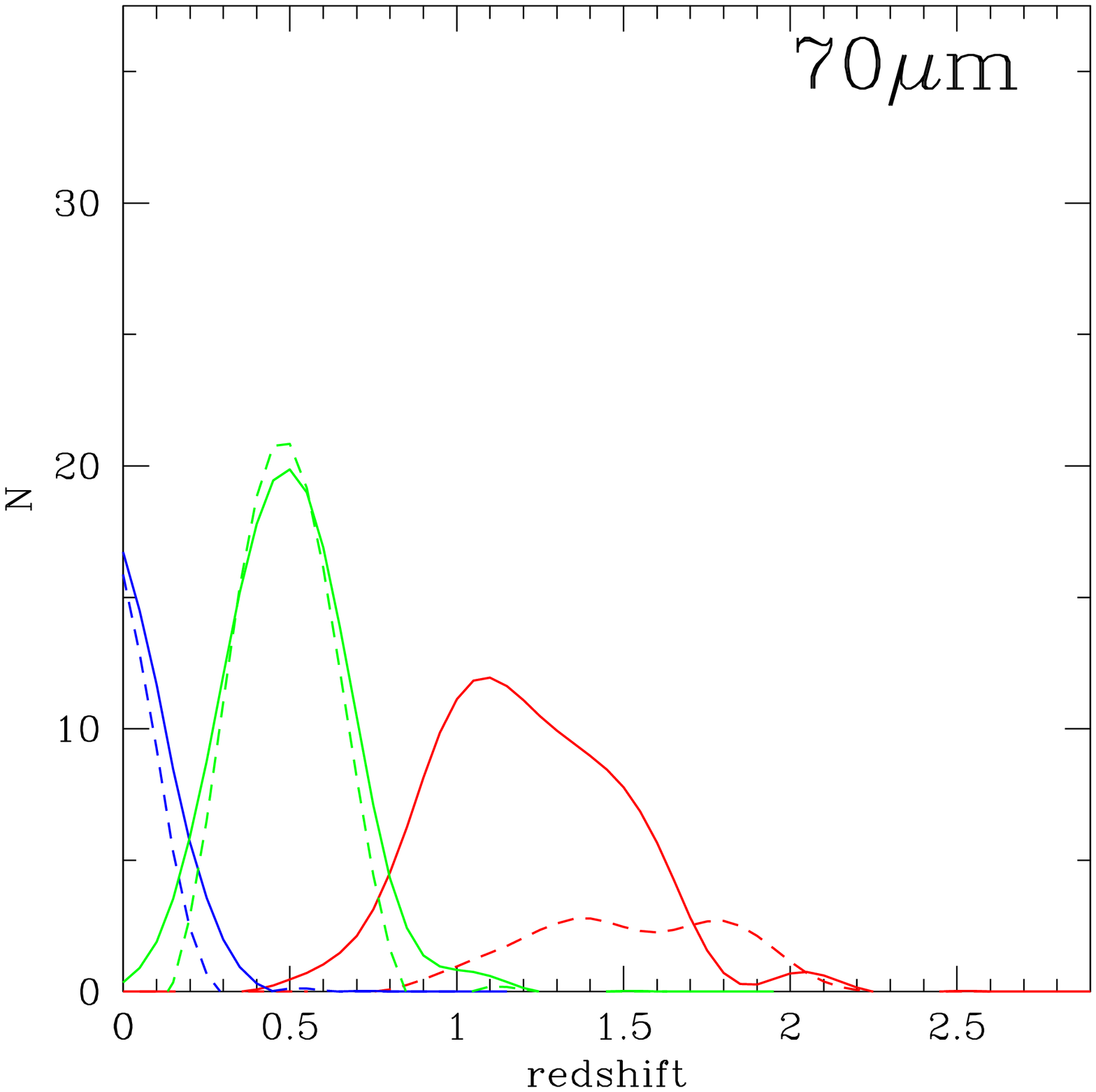,width=1.4in}
\psfig{file=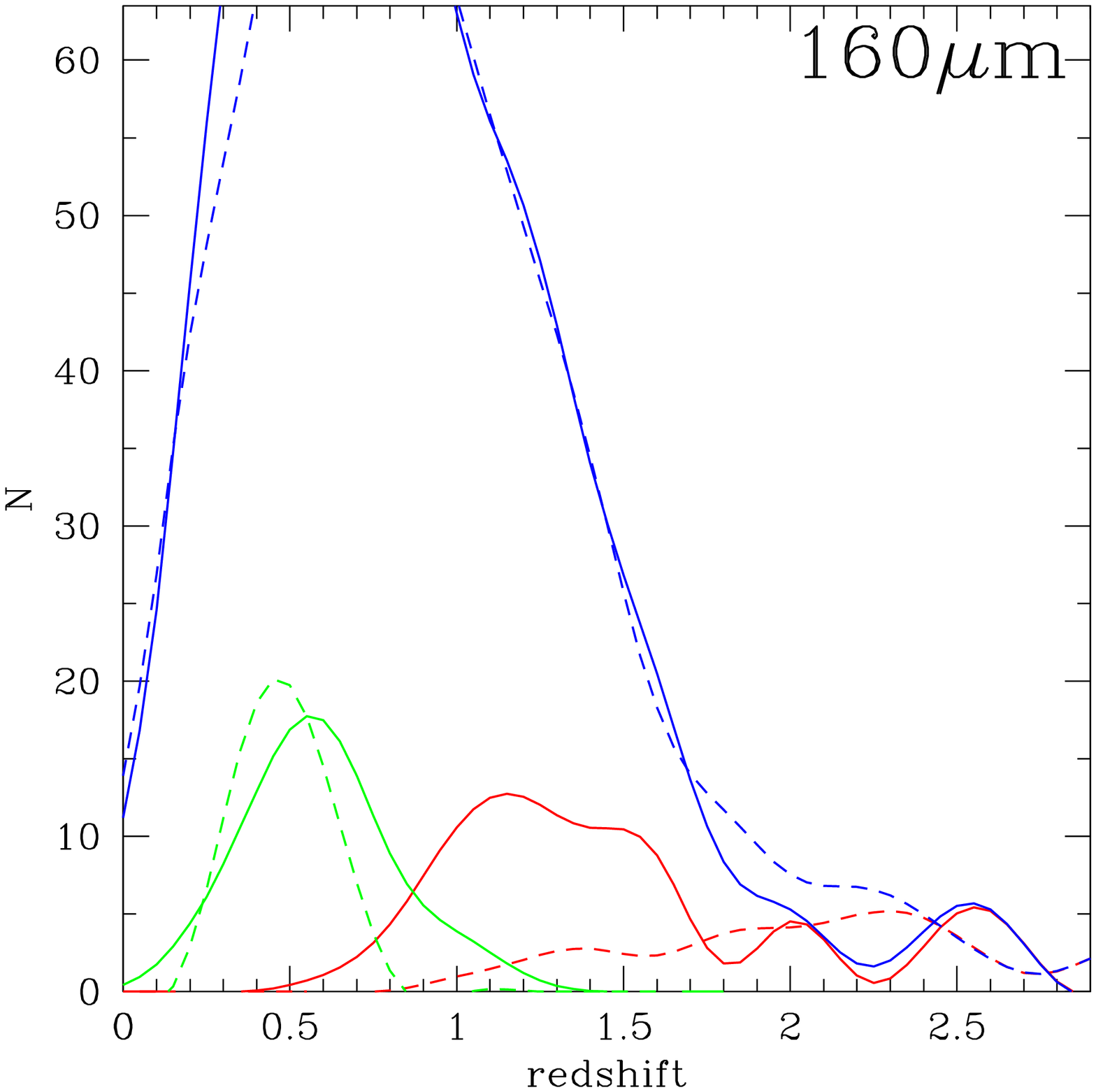,width=1.4in}
\psfig{file=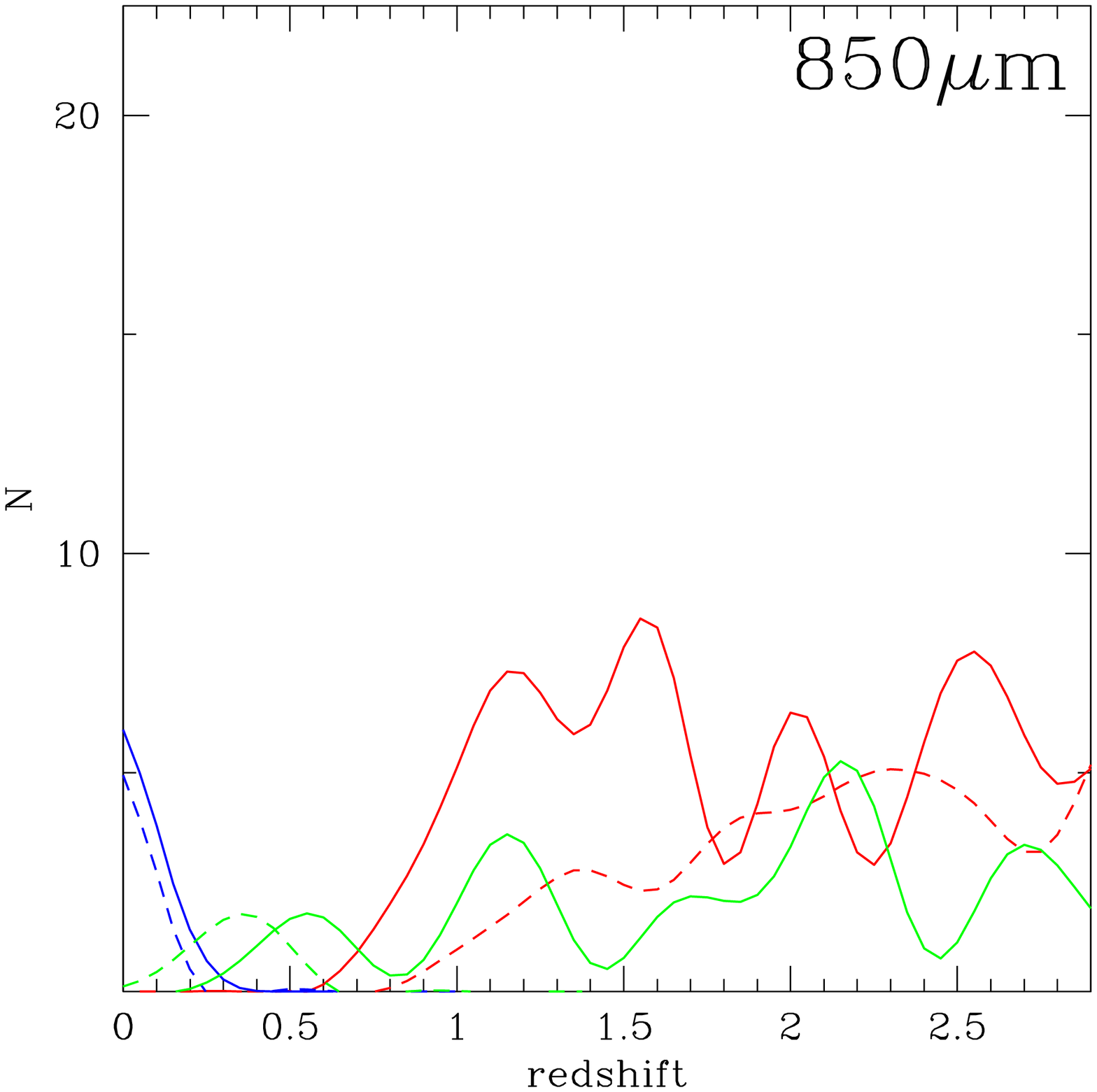,width=1.4in}
}
\caption{  Comparison  of  the  dust temperature  distributions  as  a
function  of redshift  for the  \bivar\ (solid  lines) to  an evolving
model  LF with  a one-to-one  mapping of  luminosity to  color (dashed
lines).  Distributions  are shown for hot (red  lines -- \td$>$40\,K),
medium  (green lines  -- 28$<$\td$<$33\,K),  and cold  (blue  lines --
\td$<$25\,K).  The panels are divided by wavelength, as indicated.
\label{ntd}
}
\end{figure*}

\section{Predictions for spitzer and submillimeter galaxies}

\subsection{Redshift \& color-color distributions}
Taking  our best-fit  evolution functions  for both  the  bivariate and
single variable  models, we predict redshift  distributions for galaxy
surveys at 24\mum,  70\mum, 160\mum, and 850\mum, covering  an area of
one  square   degree.   We  place  flux  density   thresholds  on  our
simulation, comparable  with a shallow {\it  Spitzer} observation such
as       the      {\it       First       Look      Survey}       ({\tt
http://ssc.spitzer.caltech.edu/fls}).  A deep survey is also analyzed,
representing flux densities five  times fainter in all bands (ignoring
the effects of confusion in the longer wavelength beams for now).  For
the submm surveys, the shallow survey is set at the confusion limit of
SCUBA  (2\,mJy). The  survey reaching  five times  this depth  will be
achievable with future instruments such as ALMA.

The comparisons are shown in Fig.~\ref{nzsirtf}.  Due to the spread in
temperature at each luminosity, with increasing numbers of hot sources
lying  at higher  redshifts  and conversely  colder  sources at  lower
redshift, the effect  of the bivariate distribution is  to increase the
numbers of both  high and low-redshift sources relative  to the single
variable  model.   In   considering  photometric  redshifts  for  {\it
Spitzer}   sources,   or    indeed   simply   trying   to   pre-select
sub-populations  of  {\it  Spitzer}  sources  with a  given  range  in
redshift or temperature properties,  it is, therefore, crucial to have
a  calibrated estimate  of  the source  distributions  in the  various
color-color planes.

\begin{figure}
\centerline{
\psfig{file=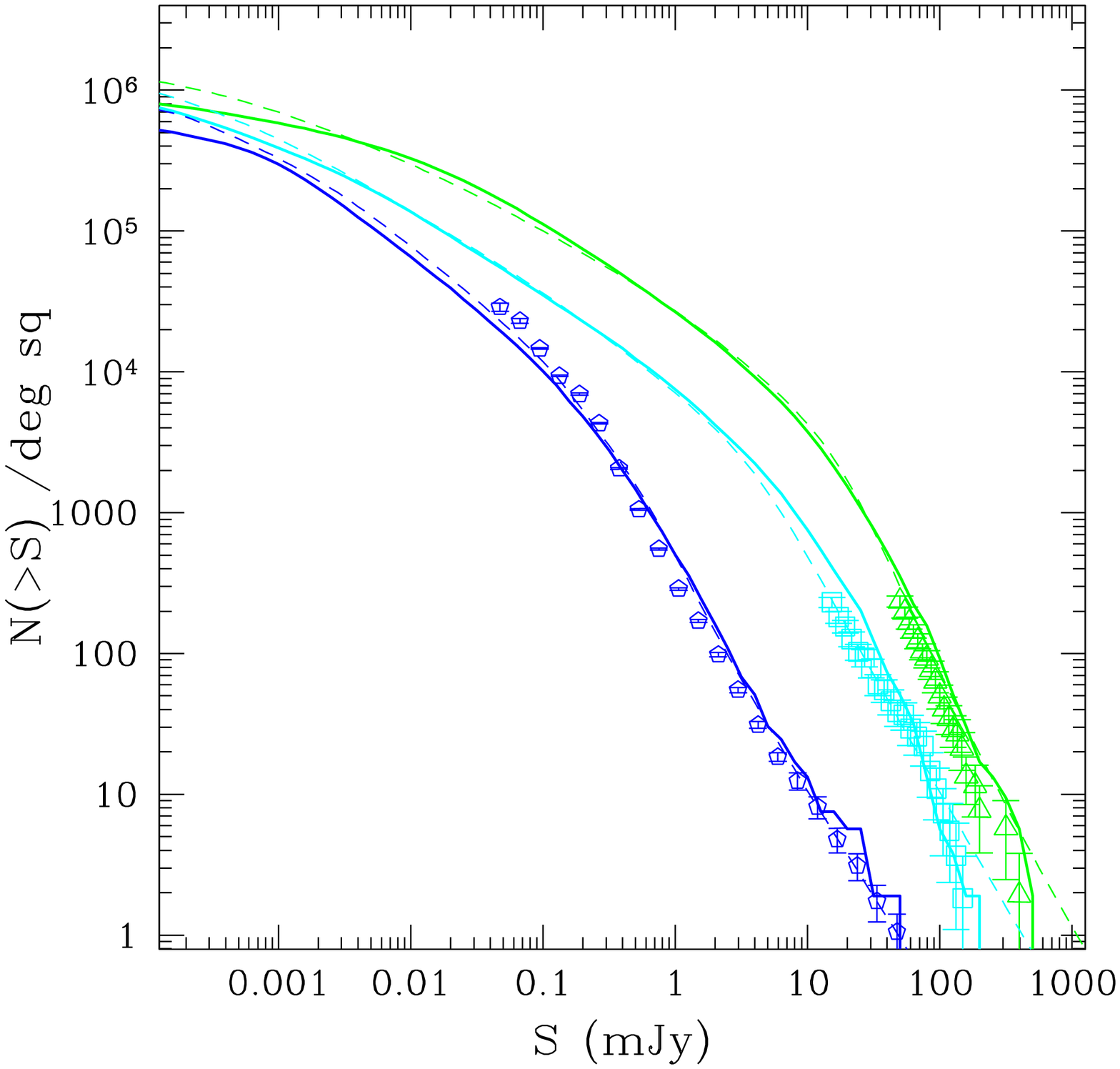,width=3.4in}
}
\caption{As for  the upper panel  in Figure~\ref{counts}, but  now the
bivariate  model is compared  to the  recently compiled  {\it Spitzer}
counts at  24$\mu m$  (blue), 70$\mu m$  (cyan) \& 160$\mu  m$ (green)
(Dole et al.  2004; Papovich et al. 2004).
\label{spitcount}}
\end{figure}

Color-color  diagrams of  galaxies in  a 24\mum\  flux  limited survey
(S$_{24  \mu   m}>0.2$\,mJy)  are  shown   in  Fig.~\ref{figcc}.   The
left-hand  panel presents  the distribution  of galaxies  assuming the
univariate  luminosity-temperature  relation,  whereas the  right-hand
panel presents the corresponding  distribution for the bivariate case.
In each  panel, objects are  color-coded by redshift:  $0<z<1$ (cyan),
$1<z<2$ (blue), $2<z<3$ (green), $z>3$ (red).  It is apparent that the
distribution  of galaxies in  the two  figures is  markedly different,
with the introduction  of a bivariate form of  the luminosity function
changing the  relative density of  galaxies in the  color-color plane;
this behavior is also seen in other color-color planes. Clearly, while
different redshift regimes can be delineated in the color-color plane,
allowing the determination of  photometric redshifts for galaxies, the
redistribution of galaxies in the plane due to the introduction of the
bivariate distribution implies that  such determinations depend on the
nature of the luminosity-temperature relationship.

While  our   model  has   not  yet  been   tuned  to  fit   the  deep,
multi-wavelength {\it Spitzer} counts,  it already emphasizes that the
source  distributions as  a  function  of redshift  and  SED can  vary
significantly,  from  the bivariate  case  to  the single  temperature
luminosity  models.   Upcoming papers  will  explore  this issue  more
fully, once deep {\it Spitzer}  data is available to further constrain
our models.

\begin{figure}
\centerline{
\psfig{file=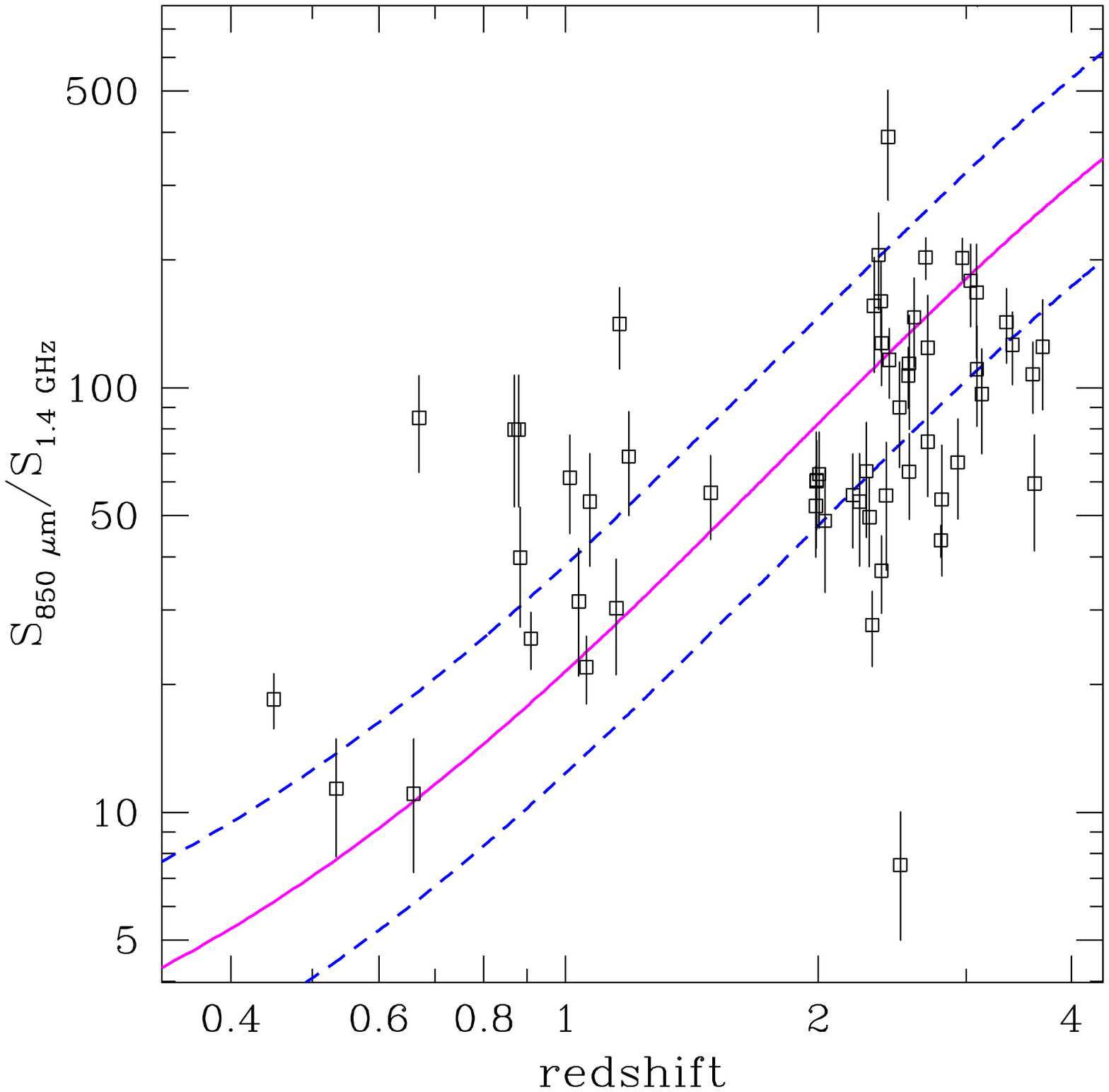,width=3.4in}
}
\caption{  \ratio indices  for SMGs  with redshifts,  compared  to our
model  prediction, with  the dashed  lines corresponding  to 1$\sigma$
scatter  in the  population of  model galaxies.   Note that  the model
includes only the  range of dust temperatures observed  locally in the
IRAS  population, whereas  a variation  in  the radio  to farIR  ratio
likely increases the scatter further.\label{figureone}}
\end{figure}

In C03 we  presented the baseline assumption in  this model: the local
\ratio\ as  a function  of ${\cal L}$  holds at all  redshifts, simply
extrapolating the local ${\cal L} - {\cal C}$ relation to the required
luminosities.  We have  now fit this model explicitly  to the pre-{\it
Spitzer}  submm and  mid-infrared  counts to  constrain the  evolution
function,  and  used this  tuned  model  to  make predictions  on  the
populations observed in the  {\it Spitzer} wavelengths (8\mum, 24\mum,
70\mum, and 160\mum).

In Fig.~\ref{lz},  we show  the \ltir\ distribution  as a  function of
redshift. Each pair of panels represents a particular waveband (8\mum,
24\mum, 70\mum,  and 160\mum\ respectively), with  the left-hand panel
in  a pair  representing the  bivariate model  (blue markers)  and the
right-hand panel presenting the  univariate case (red markers).  Until
we  apply a  flux limit  to the  figure (darker  points), there  is no
difference  between  the  visualizations  since  they  have  the  same
luminosity  evolution  formalism.  Fig.~\ref{lz}  is  our Monte  Carlo
representation  of the evolving  LF; vertical  slices reveal  the dual
power law \bivar\ at each redshift.

When  flux   limited  surveys  are   considered  in  the   context  of
Fig.~\ref{lz}, differences in the  two models become manifest.  In the
case  of the  single  variable distribution,  each  \ltir\ point  maps
uniquely to a  flux for a given wavelength.   However in the bivariate
LF,  each \ltir\ point  corresponds to  a probability  distribution of
fluxes corresponding to the log-normal distribution in \ratio\ and the
associated  range of  SED templates  that can  be tied  to  the \ltir\
value.  The sensitivity limits in  the various bands are shown for the
deepest surveys with  {\it Spitzer}.  The structure in  the 8\mum\ and
24\mum\ surveys  is a  result of PAH  bands (rest  $\sim$10\mum) being
redshifted through  the {\it Spitzer} 24\mum\ filter.   At 70\mum\ and
160\mum, the SED is relatively smooth.

The effect is subtle in Fig.~\ref{lz}, as both the luminosity function
and the color distribution are scattering the observed fluxes, largely
canceling  dramatic  differences in  the  effective luminosity  limits
probed with redshift.  However,  the most important difference between
the bivariate and luminosity-only  models is apparent in Fig.~\ref{lz}:
in  the simpler $\Phi({\cal  L})$ model,  the flux  limit for  a given
wavelength  translates at  each redshift  into a  transition  range of
luminosities, within which galaxies  are or are not detected depending
on their color.  For surveys selecting sources along the hot dust side
of  the grey-body peak,  that being  the case  for all  the accessible
wavelengths of  {\it Spitzer} except 160\mum,  colder luminous sources
will   be  missed   and  hotter   low  luminosity   sources   will  be
preferentially detected.

\begin{figure*}
\centerline{      \psfig{figure=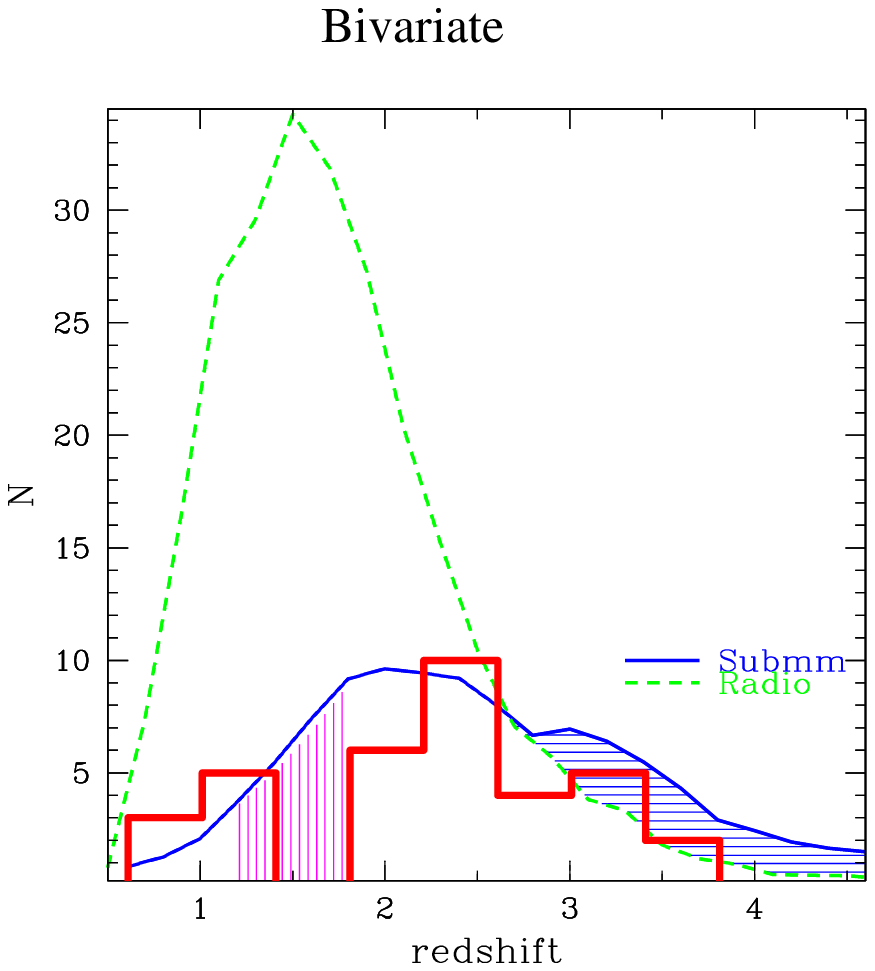,angle=270,width=2.5in}
\psfig{figure=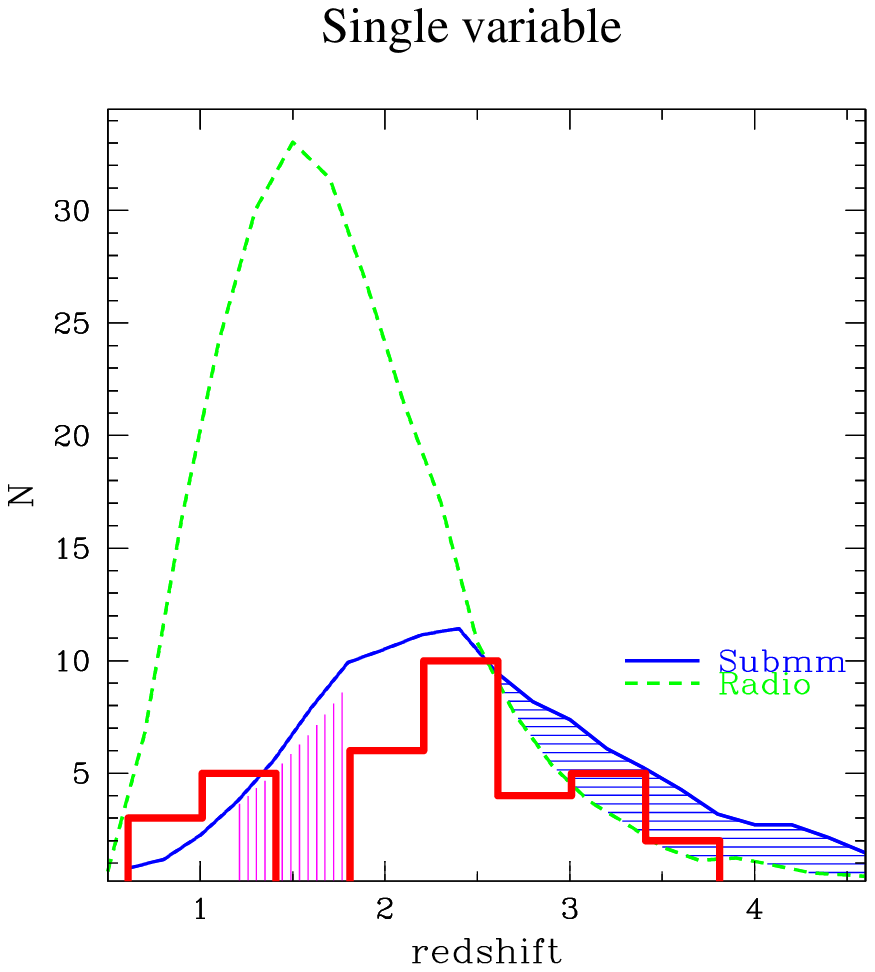,angle=270,width=2.5in} }
\caption{Predicted redshift distributions for radio (dashed) and submm
(solid)  wavelengths. The  left  panel shows  the  prediction for  the
bivariate model,  while the  right panel shows  the single  variable LF
model.  Mild differences in  the predicted N($z$) are observed between
models, which the data  cannot differentiate between.  The dark shaded
region in both panels reveals the submm sources which are not expected
to  be detectable  in the  radio.  The  light shading  shows  the {\it
spectroscopic desert} where no  strong emission lines are shifted into
the  optical  atmospheric window.   Overlaid  as  a  histogram is  the
measured redshift distribution for  radio detected submm galaxies from
Chapman et al.\ (2003c).\label{nzsmg} }
\end{figure*}

The sensitivity  limit of {\it  Spitzer} is adversely affected  by the
steep  Wien slope of  the far--mid-infrared  SED, making  more distant
sources  difficult to detect  and resulting  in the  steep sensitivity
curve in Fig.~\ref{lz}.  The result is that near the sensitivity limit
of {\it  Spitzer}, surveys will be  dominated by the  large numbers of
sources lying  both above  and below any  fixed temperature  cut.  For
hotter dust  temperatures, an excess of  a factor greater  than two of
sources is predicted  by the \bivar\ model.  All  sources are luminous
enough to be  detected regardless of their temperature,  and there are
far more lower luminosity sources which are boosted by the bivariate LF
to hotter dust temperatures than vice versa.

Fig.~\ref{ntd}   demonstrates  these  model   differences  explicitly,
demonstrating that many  more sources are detected in  the extremes of
the  temperature distributions  in  the \bivar\  model. Moreover,  the
predicted distributions  are often radically different in  form in the
\bivar\  case.  The  double peaked  profile apparent  in  the redshift
distributions  presented  in   Figure~\ref{nzsirtf}  is  a  result  of
temperature   differences    in   the   underlying    IR   population.
Fig.~\ref{ntd},  therefore,   illustrates  the  strongest  differences
between the univariate and bivariate models.

Recently deep counts at 24$\mu  m$, 70$\mu m$ and 160$\mu m$, obtained
with the MIPS intrument on  {\it Spitzer}, have become available (e.g.
Dole  et al.   2004; Papovich  et al.   2004).  This  presents further
opportunities    to    test   the    efficacy    of   the    bivariate
model. Figure~\ref{spitcount} presents  the bivariate model prediction
for the  {\it Spitzer} MIPS  bands with the observed  counts overlaid.
Generally, the  observed trends are reproduced,  but discrepancies are
apparent, most notably  a deficit in the {\it  Spitzer} counts between
1mJy -  10mJy. Addressing these  discrepancies is beyond the  scope of
this present work and will be reserved for further study.

\section{Studying the submillimeter galaxies}
{\it Spitzer} data is just  beginning to be transmitted back to Earth,
and  analysis  and  spectroscopic  followup will  take  a  significant
effort.   Detailed comparison  of our  models with  the  {\it Spitzer}
sources will be the focus of an upcoming paper, once an initial census
of the {\it Spitzer}  surveys are complete.  However our understanding
of  the  deepest  850\mum\   submm  sources  has  recently  reached  a
relatively  mature  state,   with  the  measurement  of  spectroscopic
redshifts  for a  large  sample  (Chapman et  al.\  2003b; Chapman  et
al.  {\it   in  preparation}).   In  this  section,   we  compare  our
predictions  for   the  submm   galaxies  (SMGs)  directly   with  the
measurements.

The  \ratio\  ratio provides  a  measure  of  the degenerate  quantity
(1+$z$)/T$_{\rm d}$  (Blain 1999b), coupled with any  evolution in the
Far-IR/radio  correlation.  As  the  redshift parameter  has now  been
independently  measured for  the  SMGs, we  can  directly compare  the
\ratio\ predictions  of our  model to the  SMGs.  Fig.~\ref{figureone}
shows  the \ratio  indices for  SMGs with  redshifts, compared  to our
model prediction.  Note that the model includes only the range of dust
temperatures  observed  locally  in  the IRAS  population,  whereas  a
variation in  the radio  to farIR ratio  likely increases  the scatter
further.

The actual  submm population appears to  span a larger  range than the
local  IRAS galaxy  population,  although there  are  some caveats  to
consider.   The SMG  surveys are  missing  cold sources  in the  radio
detection criterion,  and thus there  is some asymmetry in  the actual
observations  towards  hotter dust  temperatures.   At low  redshifts,
there are some apparently very  cold galaxies, which are not predicted
by  the local  IRAS  extrapolation.  If  the  SMG identifications  are
correct, they may signal a rare population which has no local analogs.

A remaining  question is the  interpretation of sources in  our model.
Ongoing  debates discuss  whether accretion  power  from super-massive
black holes or bursts of star formation is heating the dust in ULIRGs.
Our model has not assumed either explicitly. Rather, we have taken the
distribution of 60\mum/100\mum\ color found locally in the IRAS galaxy
population  and  assumed  it  describes  galaxies  at  all  redshifts.
However, AGN are known to  typically have warmer colors, and our model
may  not accurately  reflect  their possibly  larger contributions  at
higher redshift, higher luminosity.

A test for  AGN contribution is whether the  SMGs are unusually bright
in  the  X-ray, where  even  obscured  AGN  should reveal  themselves,
penetrating  high columns  of gas.   The radio  identification  of the
\sub\ sources has allowed their  location to be pinpointed.  In Barger
et al.\ (2001a,b),  it was pointed out that  approximately 10\% of the
X-ray  sources  were  \sub\  detected  in the  Chandra  Msec  exposure
centered on the HDF.  The  2\,Msec exposure in this field has detected
the majority ($\sim$ 90\%) of the radio-SMGs (Alexander et al.\ 2003).
As this radio identified \sub\ population represents $\sim$65\% of the
total blank-field  SCUBA population  at \smm$>$5\,mJy, this  implies a
fraction  $>60$\% of  the total  SCUBA population  is detectable  in a
2\,Msec Chandra  exposure.  The  typical inferred star  formation rate
from this analysis is roughly 1000$\sfr$, similar to that deduced from
the \sub\ luminosities.

The  sources with  considerably higher  X-ray luminosities,  where the
X-ray implied SFR would severely exceed the \sub\ estimate, suggest an
AGN must be  generating the X-ray emission, and  we emphasize that our
model does not account for such sources explicitly.  Note however that
in Fig.~\ref{figureone},  only one  source explicitly stands  out from
the distribution due to radio excess over the model expectations.

Let us now look specifically  at the differences between the bivariate
and single variable models.  The counts and redshift distributions are
natural  measurable  outputs from  our  model.   The univariate  model
predicts a  one-to-one mapping of  \ratio\ with redshift [for  a fixed
Q-value (Helou  et al.  1985)];  clearly the observed scatter  seen in
Figure~\ref{figureone} cannot be explained such a simple mapping.  The
bivariate model naturally introduces a scatter into this relationship,
and  the  dashed lines  in  Figure~\ref{figureone}  correspond to  the
1$\sigma$ width of the \ratio\ distribution as a function of redshift.
It is a straw man argument to  compare a single SED model to our data,
which  provides an  unphysically  narrow range  of submm/radio  colors
(e.g., C02).  A  more realistic approach is to  constrain the width of
the color distribution of our best fitting bivariate model until there
is only a  single temperature SED associated with  a given luminosity.
This  single variable  model can  be thought  of as  mapping  the dust
temperature monotonically to the source luminosity.

We first extract the counts  from our models and compare directly with
the measured  field counts.   Both models are  able to fit  the counts
well  within  current  measurement   errors.   The  submm  counts  are
insensitive to  the width in color,  ${\cal C'}$, in our  model due to
the degeneracy  in T$_{\rm d}$  versus redshift. Therefore  neither of
these models  is better constrained  by the submm counts.   The far-IR
background (Puget  et al.\  1996; Fixsen et  al.\ 1998)  also provides
little  constraint on  the detailed  form of  the  high-$z$ evolution,
requiring only that the evolution  function fall off quickly enough at
high redshifts to  avoid generating too high an  energy density.  Both
models generate submm and  far-IR background values midway between the
measurements   of  Puget   and   Fixsen.   The   model  is   therefore
self-consistent  as a generalization  to the  evolution of  the entire
submm  population.   While  the   model  does  not  provide  a  unique
description of the  submm galaxies, it is physically  motivated by the
detailed properties of local and moderate redshift IRAS galaxies.

We can use  our model to predict the  redshift distributions for radio
(1.4\,GHz) and submm (850\mum) galaxies.  In Fig.~\ref{nzsmg}, we show
the prediction  for the  bivariate model (left  panel) compared  to the
single variable LF model (right panel).  The bivariate scenario marks a
difference in the predicted N($z$),  where a deficit of sources at the
redshift peak are  seen relative to the single  variable model, with a
corresponding boost to the higher redshift sources.  This behavior was
described  in  C02, where  the  far-IR/radio  distribution provides  a
similar effect.  The shaded regions  between the radio and submm model
curves  reveal  the  submm  sources  which  are  not  expected  to  be
detectable in  the radio.   In the bivariate  model, a range  of hotter
dust temperatures  for a  given luminosity lead  to a  higher redshift
range for  radio detectability.  While  not statistically significant,
the  bivariate   models  appears  to  follow   the  observed  redshift
distribution.

Understanding  of  the  full   redshift  distribution  for  the  submm
population  through  deep  optical  and millimeter  spectroscopy  will
provide a  strong test  of these models  and the applicability  of the
local  bivariate ${\cal  L}-{\cal T}$  relation to  this  high redshift
population.  The discovery of  the redshift distribution for the radio
detected submm galaxies (Chapman et al.\ 2003a,c) is shown overlaid on
our  models  in Fig.~\ref{nzsmg}.   A  remarkable  agreement is  found
between the  data and our model,  with a preference  for the flattened
distribution and expanded range of redshifts in the bivariate picture.
We conclude that  submm galaxies exhibit at least as  large a range in
dust temperatures as local IRAS galaxies, and likely a larger range.
 
\section*{Acknowledgements}
The anonymous referee is  thanked for their constructive comments.  We
gratefully acknowledge the extensive  help of K.~Witherington with the
CFHT-12k  camera archives,  and J.C.~Cuillandre  for the  creation and
support  of  the  FLIPS  data  reduction  pipeline.   GFL  thanks  the
Austalian  Nuclear  Science and  Technology  Organization (ANSTO)  for
financial support.

\end{document}